\documentclass[12pt]{article}
\usepackage{graphicx}
\usepackage{latexsym}
\begin{document}
\renewcommand{\theequation}{\thesection.\arabic{equation}}
\newcommand{\be}{\begin{equation}}
\newcommand{\ee}{\end{equation}}
\newcommand{\beqy}{\begin{eqnarray}}
\newcommand{\eeqy}{\end{eqnarray}}
\newcommand{\p}{\partial}
\newcommand{\hp}{\widehat{\p}}
\newcommand{\ov}{\overline}
\newcommand{\da}{^{\dagger}}
\newcommand{\w}{\wedge}
\newcommand{\st}{\stackrel}
\newcommand{\mb}{\mbox}
\newcommand{\mx}{\mbox}
\newcommand{\mt}{\mathtt}
\newcommand{\dt}{\mathtt{d}}
\newcommand{\al}{\alpha}
\newcommand{\bb}{\beta}
\newcommand{\ga}{\gamma}
\newcommand{\te}{\theta}
\newcommand{\Te}{\Theta}
\newcommand{\de}{\delta}
\newcommand{\et}{\tilde{e}}
\newcommand{\ze}{\xi}
\newcommand{\s}{\sigma}
\newcommand{\e}{\epsilon}
\newcommand{\om}{\omega}
\newcommand{\Om}{\Omega}
\newcommand{\la}{\lambda}
\newcommand{\La}{\Lambda}
\newcommand{\n}{\nabla}
\newcommand{\hn}{\widehat{\nabla}}
\newcommand{\hph}{\widehat{\phi}}
\newcommand{\ah}{\widehat{a}}
\newcommand{\bh}{\widehat{b}}
\newcommand{\ch}{\widehat{c}}
\newcommand{\ddh}{\widehat{d}}
\newcommand{\eh}{\widehat{e}}
\newcommand{\ph}{\widehat{p}}
\newcommand{\qh}{\widehat{q}}
\newcommand{\mh}{\widehat{m}}
\newcommand{\nh}{\widehat{n}}
\newcommand{\Dh}{\widehat{D}}
\newcommand{\stu}{\st{\textvisiblespace}}
\newcommand{\au}{\stu{a}}
\newcommand{\bu}{\stu{b}}
\newcommand{\cu}{\stu{c}}
\newcommand{\du}{\stu{d}}
\newcommand{\eu}{\stu{e}}
\newcommand{\mmu}{\stu{m}}
\newcommand{\nnu}{\stu{n}}
\newcommand{\pu}{\stu{p}}
\newcommand{\Du}{\stu{D}}
\newcommand{\sto}{\st{\circ}}
\newcommand{\as}{\st{\circ}{a}}
\newcommand{\bs}{\st{\circ}{b}}
\newcommand{\cs}{\st{\circ}{c}}
\newcommand{\ds}{\st{\circ}{d}}
\newcommand{\es}{\st{\circ}{e}}
\newcommand{\ms}{\st{\circ}{m}}
\newcommand{\ns}{\st{\circ}{n}}
\newcommand{\ps}{\st{\circ}{p}}
\newcommand{\Ds}{\st{\circ}{D}}
\newcommand{\sts}{\st{s}}
\newcommand{\sth}{\st{\heartsuit}}
\newcommand{\stp}{\st{\perp}}
\newcommand{\std}{\st{\diamondsuit}}
\newcommand{\ad}{\st{s}{a}}
\newcommand{\bd}{\st{s}{b}}
\newcommand{\cd}{\st{s}{c}}
\newcommand{\gd}{\st{s}{g}}
\newcommand{\dd}{\st{s}{d}}
\newcommand{\Dd}{\st{s}{D}}
\newcommand{\ed}{\st{s}{e}}
\newcommand{\fd}{\st{s}{f}}
\newcommand{\zd}{\st{s}{\xi}}
\newcommand{\md}{\st{s}{m}}
\newcommand{\nd}{\st{s}{n}}
\newcommand{\stc}{\st{c}}
\newcommand{\az}{\st{c}{a}}
\newcommand{\bz}{\st{c}{b}}
\newcommand{\cz}{\st{c}{c}}
\newcommand{\dz}{\st{c}{d}}
\newcommand{\Dz}{\st{c}{D}}
\newcommand{\ez}{\st{c}{e}}
\newcommand{\fz}{\st{c}{f}}
\newcommand{\nz}{\st{c}{n}}
\newcommand{\mz}{\st{c}{m}}
\newcommand{\tb}{\overline{\theta}}
\newcommand{\ti}{\widetilde}

\newcommand{\2}{\textstyle{1\over 2}}
\newcommand{\3}{\frac{1}{3}}
\newcommand{\4}{\frac{1}{4}}
\newcommand{\8}{\frac{1}{8}}
\newcommand{\6}{\frac{1}{16}}

\newcommand{\ra}{\rightarrow}
\newcommand{\Ra}{\Rightarrow}
\newcommand{\im}{\Longleftrightarrow}
\newcommand{\hs}{\hspace{5mm}}
\newcommand{\x}{\star}
\newcommand{\Delt}{\p^{\star}}

\thispagestyle{empty}
\noindent
{\bf 13th February} \hspace{\fill}{\bf McGill - 04/01} \\
{\bf 2004} \hspace{\fill}{\bf YITP-SB-03-18}
\vspace{1cm}
\begin{center}{\Large{\bf QUINTESSENCE AND INFLATION FROM \\
SYMMETRY BREAKING TRANSITION OF  
\vspace{3mm}\\
THE INTERNAL MANIFOLD}}\\
\vspace{5mm}
{\large{\bf Tirthabir Biswas\footnote{tirtho@hep.physics.mcgill.ca}}
\bf and Prashanth Jaikumar\footnote{jaikumar@hep.physics.mcgill.ca}}\\
\vspace{5mm}
{\small Physics Department, McGill University \\
3600 University Street, Montr\'eal, Canada H3A 2T8}
\end{center}

\begin{abstract}
We show that even in the simple framework of pure Kaluza-Klein
gravity the shape moduli can generate potentials supporting inflation
and/or quintessence. Using the shape moduli  as the
inflaton or quintessence-field has the additional benefit of being able to
explain symmetry breaking in a natural geometric way. A 
numerical analysis suggests that in these models it may be possible to obtain sufficient e-foldings during inflation as
well as a small cosmological constant at the current epoch (without fine tuning), while preserving the constraint coming from the fine structure
constant.
\end{abstract}

\newpage
\setcounter{page}{1}

\section{{\bf   INTRODUCTION}}

Kaluza-Klein/Supergravity theories provide an elegant way of combining (four dimensional) gravity with gauge interactions in a  geometric way through dimensional reduction  schemes (see for example \cite{kaluza} for details). One starts with  a higher dimensional space-time containing four dimensional observable or ``external'' universe along with  extra dimensions constituting  the ``internal manifold''. The latter remains unobserved essentially due to its smallness\footnote{In the brane world scenario \cite{braneworld} large extra dimensions are also possible.}. One then usually considers the vacuum to be a product of a four-dimensional vacuum manifold (Minkowski, deSitter or anti deSitter), and a  compact internal manifold with matching scalar curvature constants. Four-dimensional physics arises as fluctuations around this vacuum. For example, if one looks at the massless modes which are  important for describing low energy physics of the higher dimensional metric, then one finds a  graviton (in the four-dimensional sector of the metric) and  gauge bosons (appearing in the off-diagonal part of the metric) associated with the Killing vectors of the ``frozen'' internal manifold. The symmetries of the internal manifold translates into gauge symmetries in the observed four dimensional universe. It is natural then to suspect that when we observe a symmetry breaking in nature (Standard Model for example), we are really observing a shadow of a symmetry breaking taking place in the internal manifold;  a dynamical transition  from a more symmetric internal space (``spherical'') to a less symmetric (``squashed'') one. In this paper we show that this indeed may be the case where the internal manifold starts off with a symmetric metric, and rolls over a potential barrier (or tunnels through) to reach a  squashed state. Such a transition would obviously have its  cosmological implications, and here we perform a preliminary analysis with respect to inflation  \cite{guth,linde}  and quintessence   \cite{ratra}. We find that the symmetry breaking can take place via two kinds of transitions: (a) the squashing field can make a transition from a symmetric vacuum to a non-symmetric vacuum as in the ordinary Higgs mechanism, the rolling over phase potentially capable of generating inflation, and  (b) the squashing field can keep evolving much like a quintessence field, effecting what we call a ``quintessential transition''! The dynamics in this case  resembles the   scenario of ``quintessential inflation'' \cite{peebles}, where the rolling over phase corresponds to inflation as before, but  at late times after the transition the internal manifold keeps getting more and more squashed accompanied by quintessence, the potential energy approaching zero asymptotically. This picture departs fundamentally from the concept of a frozen internal manifold to that of a  dynamic one. The second scenario thus also suggests a possible resolution to the long-standing problem in Kaluza-Klein/Supergravity dimensional reduction schemes of a large (of the order of Planck mass) effective four-dimensional cosmological constant as it is inversely related to the  compactification radius, once the shape is fixed.

Previously geometric mechanisms of symmetry breaking have been realized by introducing additional scalar fields \cite{sobczyk}. However, we concentrate only on  pure Kaluza-Klein gravity (no extra non-geometric scalar fields) where the internal manifold is a Lie group, say $G$, and the initial isometry group $G_L\otimes G_R$  is broken down to  $G_L\otimes H_R$ \cite{TB}. This should perhaps be viewed only as a toy model to be extended to supergravity (SUGRA). It should be mentioned that  ideas of using geometry of extra dimensions to break gauge symmetry can also be found  in the context of ``dimensional reduction by isometries'' \cite{cho}, which is however fundamentally different from the Kaluza-Klein scenario. 

Within the Kaluza-Klein/SUGRA framework squashed vacuum metrics \cite{squashed}, their stability \cite{stability}, and geometric ways of breaking symmetry  \cite{okada} have also been studied for some special internal manifolds using a quantum field theoretic approach. However, in this paper we study the dynamics from a cosmological view point. As was suggested in \cite{TB} we first identify the scalar fields corresponding to the size and the shape of the internal manifold that are relevant to study the phase transition. We obtain an effective action of these scalar fields coupled to four dimensional gravity and show that the truncation is consistent \cite{duff}, i.e. the solutions of the field equations derived from the effective action are also solutions of the complete higher dimensional Einstein's equations. One can then derive a quantum mechanical action by treating these  fields  as collective coordinates (the ``radii'' of our observational and internal dimensions, $A(t)$ and $S(t)$ respectively, and a squashing variable, $T(t)$) characterizing the internal and the external manifold.

To study the dynamics comprehensively is a difficult task but one can get significant insight by looking at the  equations of motion (for the collective coordinates), effective potentials and approximate solutions. In particular we find solutions that can be associated with the inflationary and quintessence phase. We also note that in the quintessence solution, a combination of the shape and the size remains fixed which is not surprising when the potential is a  sum of exponentials \cite{Anupam}, which is essentially what we have. Hence motivated, we make a simplifying assumption that the moduli is partially stabilized. This can also  be  achieved by several other mechanisms, like by turning on the fluxes~\cite{moduli}, wrapping branes~\cite{brane,easson,kaya} etc., at least approximately, within a given cosmological era. Technically, this assumption simplifies the analysis greatly as single scalar field potentials have been studied extensively both in the context of inflation and quintessence. In order to perform further cosmological analysis it is convenient to perform conformal rescalings of the effective field theoretic action. The scalar potential that we thus obtain for the squashing field is a sum of four exponential terms. We note here that  exponential potentials and their combinations have previously been studied both in the context of inflation~\cite{maeda} and quintessence~(\cite{copeland} and refs. therein). Depending upon the values and signs of the parameters of our potential, several interesting cases emerge, of which we mention three at this point. 

\vskip0.1cm

Firstly, for a range of parameters one can find a double well potential indicating an (a) type symmetry breaking which is also suitable for inflation; for some typical parameter values we obtained around 50 e-foldings.  For a different choice of parameters when the higher dimensional cosmological constant is set to zero, quite remarkably we find that  the potential obtained resembles the one recently discussed in~\cite{alpha} which successfully relates the evolution of an oscillating quintessence field with astrophysical data on the variation of the fine structure constant. Indeed, in our model the fine structure constant corresponding to the Kaluza-Klein gauge fields depends on  the evolving scalar fields.  

\vskip0.1cm

In our opinion however, a more interesting case is when one can realize a symmetry breaking of type (b) in a quintessential-inflation scenario. This can be achieved in two ways, one by trying to combine the two scenarios discussed above, and the other by considering a scalar field slowly rolling towards infinity. Here we mainly focus on this latter possibility and show that indeed one can realize an inflationary phase, followed by a period of radiation domination, and finally a quintessential acceleration phase, with a cosmological constant energy density
\be
\lambda=10^{-123}M_p^4\ ,\label{cosmos}
\ee
 where the Planck mass $M_p=1.2\times 10^{19}$ GeV. Further, numerical results in this case are consistent with the value of the fine structure constant and we find the masses of the broken gauge bosons to correspond to the (S)GUT scale. We emphasize that unlike previous attempts at obtaining quintessence models which used extra dimensions whose size could vary~\cite{size}, in our case it is the shape which plays the more dominant role, although we do not discount the possibility that once Supergravity/String theory effects (like branes and fluxes) are included both size and shape may become important. Also, it seems possible to embed this model in the brane-world framework making it phenomenologically more attractive\footnote{In~\cite{bshape} such a scenario has been studied, when the internal manifold is a flat tori, whereas the novel geometric effects that we obtain originate from the internal curvature of the extra dimensions.}, although its direct connection to symmetry breaking and gauge theories would be compromised.

So, could it be that the  geometric symmetry breaking mechanism in Kaluza-Klein theories can also explain inflation and quintessence? 

A conclusive answer cannot yet be provided. One has to incorporate  matter-radiation in the picture and carry out  a  more rigorous  analysis addressing issues like  primordial density fluctuations, baryogenesis \cite{joyce}, nucleosynthesis \cite{liddle}, relic particle abundance \cite{relic}, gravitational waves \cite{peebles} etc. Our preliminary estimates suggest that some fine-tuning may be neccessary to account for the observed spectral tilt and amplitude of cosmic microwave background (CMB) fluctuations. In the quintessence scenarios, since the squashing field couples to radiation in our model this leads to a time varying fine structure constant as well as acting as a fifth force in its quintessence phase. Observational bounds on time variation of $\al$ and fifth force experiments gives similar bounds on the ``effective coupling exponent'' of the squashing field to the radiation. The bounds do seem to be consistent with the quintessential inflation picture but points at the neccessity of stabilization mechanisms which freezes a specific linear combination of the size and the shape moduli. Ideally one should incorporate the moduli stabilizing effect in our analysis which  will tell us the combination of the moduli fields that are frozen in the different cosmological eras.  

This paper is organized as follows: In section 1, we introduce our model including a  brief review of group theory and identify the relevant scalar fields  in the higher dimensional metric. In section 2, we first obtain an effective action through dimensional reduction and then check the consistency of this action. In section 3, we  obtain a quantum mechanical action and equations of motion  involving the shape and the size of our universe. We proceed to obtain symmetry breaking solutions resembling inflation and quintessence. In section 4, we study the cosmological implications toward inflation and quintessence in more detail. Finally, we conclude with a brief summary and some remarks about future research directions.   

\section{{\bf  OUR MODEL }}
As in \cite{TB} we consider our universe to be a semi-direct product, $M_{D+1}\otimes G$, where $M_{D+1}$ is the $D+1$-dimensional observational universe and $G$, a Lie group manifold, serves as the Kaluza-Klein internal space \cite{kerner}. Before we decide on an ``ansatz'' for the dimensional reduction, let us quickly review the Lie group geometry. 
\vspace{5mm}
\\
{\bf Geometry of Lie groups:}  A Lie group element $g$ can be parameterized as 
\begin{equation}
g=exp(\chi^{\as}(y^{\ms})T_{\as})\ \in\ G
\end{equation}
where $T_{\as}\ \in\ \cal{G}$, the Lie algebra corresponding to the Lie group $G$ and $\chi^{\as}(y^{\ms})$ are some given functions of the coordinates $y^{\ms}$ charting the Lie group manifold. The Lie group generators $T_{\as}$ satisfy the usual commutation relations:
\begin{equation}
[T_{\as},T_{\bs}]=C_{\as\bs}{}^{\cs}T_{\cs}
\end{equation}
where $C_{\as\bs}{}^{\cs}$ are the structure constants of the Lie group. With each of the generators $T_{\as}$, one can associate a left and a right invariant vector field $e_{\as}$ and $\et_{\as}$ respectively. Both sets $\{e_{\as}\}$ and $\{\et_{\as}\}$ can serve as vielbeins or local basis vector fields for the tangent space of the Lie group. They are defined via the following relations
\begin{equation}
e_{\as}\equiv e_{\as}{}^{\ms}\p_{\ms};\ \et_{\as}\equiv \et_{\as}{}^{\ms}\p_{\ms}
\end{equation}
\begin{equation}
e_{\as}{}^{\ms}\equiv (e_{\ms}{}^{\as})^{-1};\ \et_{\as}{}^{\ms}\equiv (\et_{\ms}{}^{\as})^{-1}
\end{equation}
and
\begin{equation}
g^{-1}\p_{\ms}g=e_{\ms}{}^{\as}T_{\as} ;\ (\p_{\ms}g)g^{-1}=\et_{\ms}{}^{\as}T_{\as}
\end{equation}
These two reference frames are related by a local Lorentz transformation
\begin{equation}
\et_{\as}=D_{\as}{}^{\bs}(g)e_{\bs}
\end{equation}
where $D_{\as}{}^{\bs}(g)$ is the adjoint representation of $G$. In the subsequent discussion we will choose $\{e_{\as}\}$ as the local frame of reference. In this frame, a general metric on $G$ looks like
$$ g_{\as\bs}=g_{\as\bs}(y^{\ms})$$
However, we are interested in metrics with special symmetry properties. It can be shown that in general the isometry group of a metric will be $K_L\otimes H_R$, where $H,K\subseteq G$. In particular 
\begin{equation}
K=G \Ra g_{\as\bs}=\mbox{constants}
\end{equation}
We will be principally concerned with such left invariant metrics. These metrics are invariant under the right invariant vector fields $\{\et_{\as}\}$, but not in general under the left invariant vector fields. This follows readily from the commutation relations between them:
\begin{equation}
[e_{\as},e_{\bs}]=C_{\as\bs}{}^{\cs}e_{\cs}\ ;\ [\et_{\as},\et_{\bs}]=-C_{\as\bs}{}^{\cs}\et_{\cs}\ ;\ [\et_{\as},e_{\bs}]=0
\end{equation}
If we want the metric to be further invariant under say $H_R$, then it has to satisfy
\begin{equation}
g_{\as\bs}=D_{\as}{}^{\cs}(h)D_{\bs}{}^{\ds}(h)g_{\cs\ds}\ \forall\  h\ \in \ H
\end{equation}
If we suitably choose our generators $\{T_{\as}\}=\{T_{\az},T_{\ad}\}$ such that $\{T_{\ad}\}$ span $\cal{H}$, then the Killing vectors of this $G_L\otimes H_R$ (left) invariant metric will be the $\{\et_{\as}\}$'s and the $\{e_{\ad}\}$'s. We will always refer group quantities by a circle ($\circ$) while that of the  {\em C}oset space $G/H$ and the {\em S}ubgroup $H$ with ($c$) and  ($s$) respectively. Sometimes we may omit the symbols when it is self-evident. We will also assume the groups $G$ and $H$ to be simple and the coset decomposition to be reductive and symmetric:
\be
C_{\az\bd}{}^{\cd}=C_{\az\bz}{}^{\cz}=0
\ee 
A special case of the left invariant metric is the bivariant metric when $H=G$, i.e. it has the maximal isometry, and is invariant under both $\{\et_{\as}\}$ and $\{e_{\as}\}$'s. The Killing metric given by
\begin{equation}
g^K_{\as\bs}=-C_{\as\cs}{}^{\ds}C_{\bs\ds}{}^{\cs}
\end{equation}
is an example of such a metric. Further, the Killing metric satisfies Einstein's field equations
\begin{equation}
R_{\as\bs}=\sto{\la}g_{\as\bs}
\label{eq:4einstein}
\end{equation}
and hence is consistent with its usual identification as Kaluza-Klein vacuum,
the constant $\sto{\la}$ being referred to as the internal curvature. Contrary to this picture of an  internal manifold frozen in its maximally symmetric Killing metric, we treat it as  dynamic. In particular, we want to study whether the manifold makes a transition from the $G_L\otimes G_R$ Killing metric to a  $G_L\otimes H_R$-invariant metric, thereby effecting a gauge symmetry breaking from $G_R\ra H_R$ in four dimensions, with the broken gauge bosons associated with the $\{e_{\az}\}$'s acquiring masses as explained in \cite{TB}. The metric in this case looks like
\begin{equation}
g^S_{\as\bs}=\left( \begin{array}{cc}
g^K_{\az\bz} & 0\\
0& {\cal T}^2g^K_{\ad\bd}
\end{array} \right)
\end{equation}
where ${\cal T}^2$ is the ``squashing'' parameter. For some values of the squashing parameter, other than 1, we can also have an Einstein manifold. Thus in \cite{TB} it was suggested that the internal manifold may make a transition from say the maximally symmetric (${\cal T}=1$) Einstein space to the less symmetric one (${\cal T}\neq 1$). It is clear, what we have to do to understand this dynamics; we should treat ${\cal T}$ as a collective coordinate ${\cal T}(t)$ characterizing the shape of the group manifold. We know to have a consistent dimensional reduction ansatz one has to also include the overall size ${\cal S}(t)$ of the internal manifold. Thus our ansatz for the group metric will be given by  
\begin{equation}
g_{\as\bs}(t)={\cal S}^2(t)\left( \begin{array}{cc}
g^K_{\az\bz} & 0\\
0& {\cal T}^2(t)g^K_{\ad\bd}
\end{array} \right)
\label{eq:wmetric}
\end{equation}
{\bf Field Theoretic Ansatz:} In the Kaluza-Klein reduction scheme we now know exactly which scalars are relevant to studying the dynamics of symmetry breaking, viz. ${\cal S}(t)\ra \Psi(x)$ and ${\cal T}(t)\ra \Te(x)$. We will denote the coordinates charting the observable universe $M_{D+1}$ by $x^m$ while we use ``hatted'', $\widehat{ }$~, quantities to refer to objects corresponding to the full higher dimensional manifold. Thus $x^{\mh}$ will be used to collectively  denote $\{x^m,y^{\ms}\}$. 

Although an expression of the metric of the form (\ref{eq:wmetric}) is physically clarifying, technically it is more convenient to include the scalars in the vielbein. We choose to  parameterize the group element as 
$$g=exp(\chi^{\az}(y^{\mz})T_{\az}) exp(\chi^{\ad}(y^{\md})T_{\ad})$$ The ansatz for the full higher dimensional vielbein is then given by
\begin{equation}
\eh_{\mh}{}^{\ah}=\left( \begin{array}{ccc}
e_m{}^{a}(x) & 0&0\\
0 &\Psi(x)\es_{\mz}{}^{\az}(y)&\Psi(x)\Te(x)\es_{\mz}{}^{\ad}(y)\\
0&0& \Psi(x)\Te(x)\es_{\md}{}^{\ad}(y)
\end{array} \right)
\label{eq:4s-inv}
\ee
and
\be
\eh_{\ah}{}^{\mh}=\left( \begin{array}{ccc}
e_{a}{}^{m}(x) & 0&0\\
0 &\ \Psi^{-1}(x)\es_{\az}{}^{\mz}(y)&\Psi^{-1}(x)\es_{\az}{}^{\md}(y)\\
0&0& \Psi^{-1}(x)\Te^{-1}(x)\es_{\ad}{}^{\md}(y)
\end{array} \right)
\label{eq:4s-vielbein}
\end{equation}
The ``flat-metric'' is then just a constant
\begin{equation}
\widehat{g}_{\ah\bh}=\left( \begin{array}{cc}
g_{ab} & 0\\
0 & g^K_{\as\bs}
\end{array} \right)
\label{eq:4s-metric}
\end{equation}
We did not include the vectors in the ansatz (\ref{eq:4s-inv}-\ref{eq:4s-metric}) because we are only interested in the vacuum dynamics and the vectors appear as fluctuations around the vacuum metric.
\vspace{5mm}
\\ 
\setcounter{equation}{0}
\section{ {\bf CONSISTENT DIMENSIONAL REDUCTION }}
{\bf Effective Action:} Our next task is to obtain an effective action for the ansatz (\ref{eq:4s-inv}-\ref{eq:4s-metric}) via dimensional reduction of the higher dimensional gravitational action\be
\widehat{S}_{\widehat{D}}=\frac{1}{16\pi \hat{G}}\int dx^{\widehat{D}}\ \eh^{-1}\widehat{R}
\label{eq:4action}
\ee
In order to compute the scalar curvature $\widehat{R}$ we first need to compute the spin connections $\widehat{\om}^{\ah}{}_{\bh}$ which are uniquely defined via
\be
\ddh \widehat{\om}^{\ah}+\widehat{\om}^{\ah}{}_{\bh}\w\widehat{\om}^{\bh}=0
\label{eq:4connection}
\ee
where $\widehat{\om}^{\ah}$ are the basis 1-forms
\be
\widehat{\om}^{\ah}=dx^{\mh}e_{\mh}{}^{\ah}
\ee
For (\ref{eq:4s-inv}) the 1-forms are given by
$$\widehat{\om}^{a}=\om^a$$
$$\widehat{\om}^{\az}=\Psi\om^{\az}$$
\be
\widehat{\om}^{\ad}=\Psi\Te\om^{\ad}
\ee
With a little algebraic manipulation and guess work one can obtain the connections satisfying (\ref{eq:4connection}):
$$\widehat{\om}^{a}{}_{b}=\om^{a}{}_{b}$$
$$\widehat{\om}^{\az}{}_{b}=(e_b\Psi)\om^{\az}$$
$$\widehat{\om}^{\ad}{}_{b}=(e_b\Psi\Te)\om^{\ad}$$
$$\widehat{\om}^{\az}{}_{\bz}=\om^{\az}{}_{\bz}+(\Te^2-1)\om^{\az}{}_{\cd\bz}\om^{\cd}$$
$$\widehat{\om}^{\az}{}_{\bd}=\Te\om^{\az}{}_{\bd}$$
\be
\widehat{\om}^{\ad}{}_{\bd}=\om^{\ad}{}_{\bd}
\ee
$e_b$ is the vielbein vector
$$e_b=e_b{}^m\p_m$$
and $\om^{\as}{}_{\bs\cs}$'s are the group connection co-efficients defined by
\be
\om^{\as}{}_{\bs\cs}=g^{\as\as'}\om_{\as'\bs\cs}\ ;\ \om_{\as\bs\cs} =\2(C_{\as\bs\cs}+C_{[\as\cs\bs]})
\ee
Our next step is to evaluate the curvature 2-forms
\be
\widehat{\cal{R}}^{\ah}{}_{\bh}=\ddh \widehat{\om}^{\ah}{}_{\bh}+\widehat{\om}^{\ah}{}_{\ch}\w\widehat{\om}^{\ch}{}_{\bh}
\ee
A straight forward computation yields the following results:
$$\widehat{\cal{R}}^{a}{}_{b}=\mathcal{R}^{a}{}_{b}$$
$$\widehat{\cal{R}}^{\az}{}_{b}=\Psi^{-1}\n_c(e_b\Psi)\widehat{\om}^{c}\w\widehat{\om}^{\az}+\Psi^{-1}(e_b\Te)\om^{\az}{}_{\cd\dz}\widehat{\om}^{\dz}\w\widehat{\om}^{\cd}$$
$$\widehat{\cal{R}}^{\az}{}_{\bz}=\mathcal{R}^{\az}{}_{\bz}$$
$$+\{(\Te^2-1)[\om^{\az}{}_{\cd\dz}\om^{\dz}{}_{\bz\ed}-\om^{\dz}{}_{\cd\bz}\om^{\az}{}_{\dz\ed}+\om^{\az}{}_{\fd\bz}\om^{\fd}{}_{\cd\ed}]+(\Te^2-1)^2\om^{\az}{}_{\cd\dz}\om^{\dz}{}_{\ed\bz})\}\Psi^{-2}\Te^{-2}\widehat{\om}^{\cd}\w\widehat{\om}^{\ed}$$
$$+\{(\Te^2-1)(\om^{\az}{}_{\cd\dz}\om^{\dz}{}_{\cd\ez}+\om^{\az}{}_{\cd\bz}\om^{\cd}{}_{\dz\ez})-(\p\Psi)^2\de_{\dz}{}^{\az}g_{\ez\bz}\}\Psi^{-2}\widehat{\om}^{\dz}\w\widehat{\om}^{\ez}+2\Psi^{-2}(e_d\Te)\om^{\az}{}_{\cd\bz}\widehat{\om}^{d}\w\widehat{\om}^{\cd}$$
$$\widehat{\cal{R}}^{\ad}{}_{b}=\Psi^{-1}\Te^{-1}\n_c(e_b\Psi\Te)\widehat{\om}^{c}\w\widehat{\om}^{\ad}+\Psi^{-1}(e_b\Te)\om^{\ad}{}_{\cz\dz}\widehat{\om}^{\cz}\w\widehat{\om}^{\dz}$$
$$\widehat{\cal{R}}^{\ad}{}_{\bd}=\mathcal{R}^{\ad}{}_{\bd}-2\Psi^{-2}\Te^{-2}(\p\Te\Psi)^2\widehat{\om}^{\ad}\w\widehat{\om}_{\bd}+\Psi^{-2}(\Te^2-1)\om^{\ad}{}_{\cz\dz}\om^{\cz}{}_{\bd\ez}\widehat{\om}^{\dz}\w\widehat{\om}^{\ez}$$
$$
\widehat{\cal{R}}^{\az}{}_{\bd}=\Te\mathcal{R}^{\az}{}_{\bd}+\Psi^{-1}e_c\Te\om^{\az}{}_{\bd\dz}\widehat{\om}^{c}\w\widehat{\om}^{\dz}-\Psi^{-2}\Te^{-1}(e_c\Psi)(e^c\Te\Psi)\widehat{\om}^{\az}\w\widehat{\om}^{\bd}$$
\be
+\Psi^{-2}(\Te^2-1)\om^{\az}{}_{\dd\cz}\om^{\cz}{}_{\bd\ez}\widehat{\om}^{\dd}\w\widehat{\om}^{\ez}
\ee
The coefficients of the Riemannian tensor can now be read off  from the curvature 2-forms
\be
\widehat{\cal{R}}^{\ah}{}_{\bh}=\widehat{R}^{\ah}{}_{\bh|\ch\ddh|}\widehat{\om}^{\ch}\w\widehat{\om}^{\ddh}
\ee
Here $|\ |$ indicates that the sum counts a pair only once. The Riemannian coefficients obtained thus are
$$\widehat{R}^{a}{}_{bcd}=R^{a}{}_{bcd}$$
$$\widehat{R}^{\az}{}_{bc\dz}=\Psi^{-1}\n_c(e_b\Psi)\de_{\dz}{}^{\az}$$
$$\widehat{R}^{\az}{}_{b\cz\dd}=\Psi^{-1}e_b\Te\om^{\az}{}_{\dd\cz}$$
$$\widehat{R}^{\az}{}_{\bz\cz\dz}=\Psi^{-2}[R^{\az}{}_{\bz\cz\dz}+(\Te^2-1)(\om^{\az}{}_{\ed[\cz}\om^{\ed}{}_{\bz\dz]}+\om^{\az}{}_{\ed\bz}\om^{\ed}{}_{[\cz\dz]})-(\p\Psi)^2\de_{[\cz}{}^{\az}g_{\dz]\bz}]$$
$$\widehat{R}^{\az}{}_{\bz\cz\dd}=\Psi^{-2}\Te^{-1}R^{\az}{}_{\bz\cz\dd}$$
$$\widehat{R}^{\az}{}_{\bz\cd\dd}=\Psi^{-2}\Te^{-2}[R^{\az}{}_{\bz\cd\dd}+(\Te^2-1)(\om^{\az}{}_{[\cd\ez}\om^{\ez}{}_{\bz\dd]}-\om^{\ez}{}_{[\cd\bz}\om^{\az}{}_{\ez\dd]}+\om^{\az}{}_{\ed\bz}\om^{\ed}{}_{[\cd\dd]})$$
$$+(\Te^2-1)^2\om^{\az}{}_{[\cd\ez}\om^{\ez}{}_{\dd]\bz}]$$
$$\widehat{R}^{\az}{}_{\bz c\dd}=2\Psi^{-1}(e_c\Psi)\om^{\az}{}_{\dd\bz}$$
$$\widehat{R}^{\ad}{}_{bc\dd}=\Psi^{-1}\Te^{-1}\n_c(e_b\Psi\te)\de_{\dd}{}^{\ad}$$
$$\widehat{R}^{\ad}{}_{b \cz\dz}=\Psi^{-1}(e_b\Te)\om^{\ad}{}_{[\cz\dz]}$$
$$\widehat{R}^{\ad}{}_{\bd\cz\dz}=\Psi^{-2}[R^{\ad}{}_{\bd\cz\dz}+(\Te^2-1)\om^{\az}{}_{\ez[\cz}\om^{\ez}{}_{\bd\dz]}]$$
$$\widehat{R}^{\ad}{}_{\bd\cd\dz}=\Psi^{-2}\Te^{-1}R^{\ad}{}_{\bd\cd\dz}$$
$$\widehat{R}^{\ad}{}_{\bd\cd\dd}=\Psi^{-2}\Te^{-2}[R^{\ad}{}_{\bd\cd\dd}-(\p\Psi\Te)^2\de_{[\cd}{}^{\ad}g_{\bd\dd]}]$$
$$\widehat{R}^{\az}{}_{\bd\cz\dz}=\Psi^{-2}\Te R^{\az}{}_{\bd\cz\dz}$$
$$\widehat{R}^{\az}{}_{\bd\cd\dz}=\Psi^{-2}[R^{\az}{}_{\bd\cd\dz}+(\Te^2-1)\om^{\az}{}_{\cd\ez}\om^{\ez}{}_{\bd\dz}+\Psi^{-1}(\p\Psi)(\p\Psi\Te)\de_{\dz}{}^{\az}g_{\bd\cd}]$$
$$\widehat{R}^{\az}{}_{\bd\cd\dd}=\Psi^{-2}\Te^{-1}R^{\az}{}_{\bd\cd\dd}$$
\be
\widehat{R}^{\az}{}_{\bd c\dz}=\Psi^{-1}(e_c\Te)\om^{\az}{}_{\bd\dz}
\ee
From the Riemann tensor it is easy to calculate the Ricci tensor
\be
\widehat{R}_{\bh \ddh}=\widehat{R}^{\ah}{}_{\bh \ah\ddh}
\ee
After some simplifications one obtains
$$\widehat{R}_{a b}=R_{a b}-\Ds\Psi^{-1}\n_b(e_a\Psi)-\Dd\Te^{-1}\n_b(e_a\Te)-\Dd\Psi^{-1}\Te^{-1}e_{(a}\Psi e_{b)}\Te$$
$$\widehat{R}_{\az \bz}=g_{\az \bz}[-\{\Psi^{-1}\Box\Psi+(\Ds-1)\Psi^{-2}(\p\Psi)^2+\Dd\Psi^{-1}\Te^{-1}\p_{a}\Psi \p^{a}\Te\}+\Psi^{-2}\sto{\la}-\4\Psi^{-2}(\Te^2-1)]$$
$$
\widehat{R}_{\ad \bd}=g_{\ad \bd}[-\{\Psi^{-1}\Box\Psi+\Te^{-1}\Box\Te+(\Ds-1)\Psi^{-2}(\p\Psi)^2+(\Dd-1)\Te^{-2}(\p\Te)^2$$
\be
+(\Dd+\Ds)\Psi^{-1}\Te^{-1}\p_{a}\Psi \p^{a}\Te\}+\Psi^{-2}(\sto{\la}-k\sts{\la})+\4\Psi^{-2}(\Te^2-1)(1-k)+\Psi^{-2}\Te^{-2}k\sts{\la}]
\label{eq:4ricci}
\ee
Here we have introduced a group theoretical parameter k:
\be
\sts{g}_{\ad \bd}=k\sto{g}_{\ad \bd}
\ee
$\sts{g}_{\ad \bd}$ is the Killing metric of group $H$ while $\sto{g}_{\ad \bd}$  of course corresponds to the Killing metric of group $G$. For a symmetric coset decomposition it is known that
\be
k=1-\frac{\Dz}{2\Dd}
\ee
If, $H$ is not a simple group but it is a product of $U(1)$'s then also the  value of $k$ is known: 
\be
k=0
\ee
$\sts{\la}$ is defined in the usual way as in (\ref{eq:4einstein}) except that now all the quantities refer to the subgroup $H$. In fact for Killing metrics 
$$\sts{\la}=\sto{\la}=\4$$
and we will explicitly substitute their values.

We are ready to compute the scalar curvature that we need in the action.
\be
\widehat{R}=g^{\ah \bh}\widehat{R}_{\ah \bh}=g^{a b}\widehat{R}_{a b}+g^{\az \bz}\widehat{R}_{\az \bz}+g^{\ad \bd}\widehat{R}_{\ad \bd}
\ee
Finally, we have
$$\widehat{R}=R-\left[2\Ds\frac{\Box\Psi}{\Psi}+2\Dd\frac{\Box\Te}{\Te}+\Ds(\Ds-1)\frac{(\p\Psi)^2}{\Psi^2}+\Dd(\Dd-1)\frac{(\p\Te)^2}{\Te^{2}}+2\Dd(\Ds+1)\frac{\p_{a}\Psi \p^{a}\Te}{\Psi\Te}\right]$$
\be
+\4\left[  (\Dz+2\Dd(1-k))\frac{1}{\Psi^2}-\Dd(1-k) \frac{\Te^2}{\Psi^2}+ k\Dd \frac{1}{\Psi^2\Te^2} \right]
\label{eq:4curvature}
\ee
Since $\widehat{R}$ is independent of the group coordinates one can perform the integration over the group in the action (\ref{eq:4action}) which essentially just yields a volume factor $V_G$. Thus we have our effective $D+1$-dimensional action
\be
S_{\mt{grav}}=\frac{V_G}{16\pi \hat{G}}\int e^{-1}\Psi^{\stackrel{\circ}{D}}\Te^{\Dd}\widehat{R}\equiv \frac{1}{16\pi G}\int e^{-1}\Psi^{\stackrel{\circ}{D}}\Te^{\Dd}\widehat{R}
\ee
It is useful to perform some integration by parts. The simplified action looks like
\be
S_{\mt{grav}}=\frac{1}{16\pi G}\int dx^{D+1}\ e^{-1}\Psi^{\Ds}\Te^{\Dd}[R-K+V]
\ee
where we have defined the Kinetic and Potential like terms for the scalar fields as
\be
K=-\left[\Ds(\Ds-1)\frac{(\p\Psi)^2}{\Psi^2}+\Dd(\Dd-1)\frac{(\p\Te)^2}{\Te^{2}}+2\Dd(\Ds-1)\frac{\p_{a}\Psi \p^{a}\Te}{\Psi\Te}\right]
\ee
and
\be
V=\4\left[ 2\Dz\frac{1}{\Psi^{2}}-\Dz\frac{\Te^2}{2\Psi^{2}}+k\Dd\frac{1}{\Psi^{2}\Te^{2}}\right]
\ee
We have also specialized to the case when $H$ is simple. At this point it is useful to redefine the scalars:
\be
\Psi=e^{\psi};\mbox{ and }\Te=e^{\te}
\ee
The kinetic and potential terms then look like
\be
K=-\left[\Ds(\Ds-1)(\p\psi)^2+\Dd(\Dd-1)(\p\te)^2+2\Dd(\Ds-1)\p_{a}\psi \p^{a}\te\right]
\ee
and
\be
V=\4\left[ 2\Dz e^{-2\psi}-\2\Dz e^{2(\te-\psi)}+k\Dd e^{-2(\psi+\te)}\right]
\label{eq:4effpot}
\ee
The action is given by
\be
S_{\mt{grav}}=\frac{1}{16\pi G}\int dx^{D+1}\ e^{-1}e^{\Ds\psi+\Dd\te}[R-K+V]
\label{eq:4e-action}
\ee
Finally, one can also include a cosmological term in the higher dimensional action
\be
\widehat{S}_{\mt{cos}}=-\frac{2\hat{\La}}{16\pi \hat{G}}\int dx^{\widehat{D}}\ \eh^{-1}
\ee
The corresponding term in the effective action is
\be
S_{\mt{cos}}=-\frac{2\hat{\La}}{16\pi G}\int dx^{D+1}\ e^{-1}e^{\Ds\psi+\Dd\te}
\label{eq:4c-action}
\ee
\vspace{5mm}
\\ 
{\bf Consistency of the Truncation:} Having obtained the dimensionally reduced field theoretic action for our model it  is time to  check the consistency of our ansatz \cite{duff}. We have to check that the solutions that we obtain by varying the effective action (\ref{eq:4e-action}) are indeed solutions of the full higher dimensional Einstein's equations, and this would mean that the truncation we performed is legitimate. 

To obtain Einstein's field equations we essentially have to compute the Einstein tensor
\be
\widehat{G}_{\ah\bh}=\widehat{R}_{\ah\bh}-\2\widehat{R}\widehat{g}_{\ah\bh}
\ee
Using (\ref{eq:4ricci}) and (\ref{eq:4curvature}) we obtain
\be
\widehat{G}_{ab}=\widehat{R}_{ab}-\2 g_{ab}\widehat{R}
\label{eq:4Gab}
\ee
$$
\widehat{G}_{\az\bz}=\sto{g}_{\az\bz}\left[(\Ds-1)\Psi^{-1}\Box\Psi+\Dd\Te^{-1}\Box\Te+\2(\Ds-1)(\Ds-2)\Psi^{-2}(\p\Psi)^2 \right.$$
$$+\2\Dd(\Dd-1)\Te^{-2}(\p\Te)^2+\Dd\Ds\Psi^{-1}\Te^{-1}\p_{a}\Psi \p^{a}\Te$$
\be
\left.+\Psi^{-2}\4(2-\Dz)-\8\Dd k\Psi^{-2}\Te^{-2}+\6(\Dz-4)\Psi^{-2}\Te^{2}-\2 R\right]
\label{eq:4Gabc}
\ee
and
$$
\widehat{G}_{\ad\bd}=\sto{g}_{\ad\bd}\left[(\Ds-1)\Psi^{-1}\Box\Psi+(\Dd-1)\Te^{-1}\Box\Te+\2(\Ds-1)(\Ds-2)\Psi^{-2}(\p\Psi)^2 \right.$$
$$+\2(\Dd-1)(\Dd-2)\Te^{-2}(\p\Te)^2+(\Dd-1)\Ds\Psi^{-1}\Te^{-1}\p_{a}\Psi \p^{a}\Te$$
\be
\left.-\4\Dz\Psi^{-2}-\8(2-\Dd) k\Psi^{-2}\Te^{-2}+\6\frac{\Dz}{\Dd}(\Dd+2)\Psi^{-2}\Te^{2}-\2 R\right]
\label{eq:4Gabs}
\ee
The pure gravity field equations read
\be
 \widehat{G}_{\ah\bh}=0
\ee
Our task is to show that the field equations that one obtains by varying the effective action (\ref{eq:4e-action}) also satisfies (\ref{eq:4Gab})-(\ref{eq:4Gabs}). 

Since 
$$\widehat{g}^{mn}=g^{mn}$$
i.e. there has been no field redefinition involving the four dimensional part of the metric, it is obvious that 
$$\frac{\de S_{\mt{grav}}}{\de g^{mn}}=  \widehat{G}_{mn}$$
\be
\Ra \frac{\de S_{\mt{grav}}}{\de g^{mn}}=0 \Longleftrightarrow \widehat{G}_{ab}=0
\ee
Thus we are left to show that 
$$\{\frac{\de S_{\mt{grav}}}{\de \Psi}=0,\frac{\de S_{\mt{grav}}}{\de \Te}=0\}\equiv \{\widehat{G}_{\as\bs}=0\}$$
A straight forward computation yields the field equations
$$\frac{\de S_{\mt{grav}}}{\de \Psi}=\frac{1}{16\pi G}e^{-1}\Psi^{\Ds-1}\Te^{\Dd}\left\{\Ds R-2\Ds(\Ds-1)\Psi^{-1}\Box\Psi-2\Dd(\Ds-1)\Te^{-1}\Box\Te \right.$$
$$-\Ds(\Ds-1)(\Ds-2)\Psi^{-2}(\p\Psi)^2 
-\Dd(\Dd-1)(\Ds-2)\Te^{-2}(\p\Te)^2-2\Dd\Ds(\Ds-1)\Psi^{-1}\Te^{-1}\p_{a}\Psi \p^{a}\Te$$
\be
\left. +\4(\Ds-2)\Psi^{-2}\left[2\Dz+\Dd k\Te^{-2}-\2\Dz\Te^{2} \right]\right\} =0
\label{eq:4psi}
\ee
and 
$$
\frac{\de S_{\mt{grav}}}{\de \Te}=\frac{1}{16\pi G}e^{-1}\Psi^{\Ds}\Te^{\Dd-1}\left\{\Dd R-2\Dd(\Ds-1)\Psi^{-1}\Box\Psi-2\Dd(\Dd-1)\Te^{-1}\Box\Te \right. $$
$$-\Dd(\Ds-1)(\Ds-2)\Psi^{-2}(\p\Psi)^2 
-\Dd(\Dd-2)(\Dd-1)\Te^{-2}(\p\Te)^2-2\Dd(\Dd-1)\Ds\Psi^{-1}\Te^{-1}\p_{a}\Psi \p^{a}\Te$$
\be
\left.+\Psi^{-2}\4\left[(2\Dz\Dd+\Dd k(\Dd-2)\Te^{-2}-\2\Dz(\Dd+2)\Te^{2}\} \right]\right\} =0
\label{eq:4theta}
\ee
Subtracting (\ref{eq:4theta}) from (\ref{eq:4psi}) gives us
$$\Dz\left[R-2(\Ds-1)\Psi^{-1}\Box\Psi-2\Dd\Te^{-1}\Box\Te-(\Ds-1)(\Ds-2)\Psi^{-2}(\p\Psi)^2 \right.$$
$$-\Dd(\Dd-1)\Te^{-2}(\p\Te)^2-2\Dd\Ds\Psi^{-1}\Te^{-1}\p_{a}\Psi \p^{a}\Te$$
$$
\left.-\2\Psi^{-2}(2-\Dz)+\4\Dd k\Psi^{-2}\Te^{-2}-\8(\Dz-4)\Psi^{-2}\Te^{2}\right]=0
$$
$$\Ra \widehat{G}_{\az\bz}=0$$
Also, by inspection 
$$(\ref{eq:4theta})\Ra \widehat{G}_{\ad\bd}= 0$$
We have thus succeeded in showing that the action (\ref{eq:4e-action}) is indeed consistent. It is easy to see that an addition of the cosmological term (\ref{eq:4c-action}) does not change the consistency of the truncation. 
\setcounter{equation}{0}
\section{ {\bf QUANTUM MECHANICS WITH COLLECTIVE COORDINATES}}
{\bf The Quantum Mechanical Action and Equations of Motion :} In the previous section we obtained the dimensionally reduced field theoretic action for our model. Our aim now is to look at some cosmological solutions for the background fields and thus we assume the fields to only depend on time. In other words we use the fields as collective coordinates characterizing the observed and the internal space-time. For the internal space we already have 
\be
\psi(x)\ra S(t) \mbox{ and } \Te(x)\ra T(t)
\ee
 characterizing the size and the shape of the internal space respectively. For the external space we draw upon the standard  cosmological picture of an expanding universe:
\be
\mt{d}s^2=-e^{2W(t)}\mt{d}t^2+e^{2A(t)}\stackrel{\textvisiblespace}{\dt}s^2
\label{eq:4replacement}
\ee
$A(t)$ is the usual cosmological radius of our universe while $W(t)$ corresponds to a gauge freedom which will be useful for later computations. We will also assume that the spatial metric $\st{\textvisiblespace}{\mt{d}}s^2$ is flat, which recent observational data seem to suggest, and use the symbol  blank space,  $\textvisiblespace$, to denote quantities corresponding to the space part of the observed space-time. Symbolically the full metric then looks like
\be
\widehat{\dt} s^2=-e^{2W(t)}\mt{d}t^2+e^{2A(t)} \stackrel{\textvisiblespace}{\dt}s^2+e^{2S(t)}( \stc{\dt}s^2+e^{2T(t)} \sts{\dt}s^2)
\label{eq:4q-metric}
\ee

To obtain a quantum mechanical action from (\ref{eq:4e-action}) we basically need to calculate $R$ for the metric (\ref{eq:4q-metric}). Again, it is useful to cast the problem in terms of the vielbein. We define
\begin{equation}
e_m{}^a= \left( \begin{array}{cc}
e^{W(t)} & 0 \\
0  & e^{A(t)}\de_{\mmu}{}^{\au} 
\end{array} \right)
\end{equation}
We can now apply the same formalism as we used to calculate $\widehat{R}$. Alternatively, we can use conformal transformation by a scale factor $exp(A)$ to obtain $R$ from $R'=0$ for the trivial vielbein
$$
e'_m{}^a= \left( \begin{array}{cc}
e^{W(t)-A(t)} & 0 \\
0  & \de_{\mmu}{}^{\au} 
\end{array} \right)
$$
In any case, one obtains 
\be
R=De^{-2W}[2\ddot{A} -2 \dot{A}\dot{W}+(D+1)\dot{A}^2]
\ee

Substituting $R$ and making the replacement (\ref{eq:4replacement})  we have the full quantum mechanical  action for the collective coordinates $W(t),A(t),S(t)$ and $T(t)$ from the effective gravitational action (\ref{eq:4e-action}).
\be
S_{\mt{qm,g}}=\int \mathtt{d} t\ e^{\Ds S+\Dd T+DA-W}[-K+V]
\label{eq:4q-action}
\ee
with
$$
K=-2D\ddot{A}-D(D+1)\dot{A}^2+\Ds(\Ds-1)\dot{S}^2+\Dd(\Dd-1)\dot{T}^2$$
\be
+2[\Dd(\Ds-1)\dot{T}\dot{S}+D\dot{A}\dot{W}]
\label{eq:4kinetic}
\ee
and
\be
V=\4e^{2(W-S)}\left[2\Dz+\Dd ke^{-2T}-\2\Dz e^{2T}\right]
\ee
It is also simple to include the cosmological term (\ref{eq:4c-action}) in the quantum mechanical  action
\be
S_{\mt{qm,c}}=-2\La\int \mathtt{d} t\ e^{\Ds S+\Dd T+DA+W}
\ee
so that the total action becomes
\be
S_{\mt{qm,eff}}=S_{\mt{qm,g}}+S_{\mt{qm,c}}
\ee

Inspection of the action (\ref{eq:4q-action}) tells us that it greatly simplifies if we choose the gauge
\be
W=DA+\Ds S+\Dd T
\ee
We no longer have a non-linear sigma model, but rather a sum of ordinary kinetic terms. One can always transform back the results to the more familiar $W=0$ gauge. A similar gauge  was recently used \cite{kaya} in the context of brane gas cosmology at the level of field equations.  In this ``canonical gauge'' the effective action becomes
\be
S_{\mt{qm}}=\int \mathtt{d} t\ [K_{\mt{qm}}-V_{\mt{qm}}]
\label{eq:4qe-action}
\ee
with
$$
K_{\mt{qm}}=D(D-1)\dot{A}^2+\Ds(\Ds-1)\dot{S}^2+\Dd(\Dd-1)\dot{T}^2$$
\be
+2[D\Ds\dot{A}\dot{S}+D\Dd\dot{A}\dot{T}+\Dd(\Ds-1)\dot{T}\dot{S}]
\ee
and
\be
V_{\mt{qm}}=\4e^{2(DA+(\Ds-1)S)}\left[2\Dz e^{2\Dd T}+\Dd ke^{2(\Dd-1)T}-\2\Dz e^{2(\Dd+1)T}\right]-2\La e^{2(\Ds S+\Dd T+DA)}
\label{eq:4q-potential}
\ee

In obtaining the effective quantum mechanical action we have ignored the total derivative terms and dropped some prefactors. To understand the dynamics we now  look at equations of motion which can be derived  by varying the action (\ref{eq:4qe-action}).
$$
\frac{\de S_{\mt{qm}}}{\de A}=0\ \Ra \ 2(D-1)\ddot{A} +2\Ds\ddot{S}+2\Dd\ddot{T}$$
$$+e^{2(DA+(\Ds-1)S+\Dd T)}\left[\Dz +\2\Dd ke^{-2T}-\4\Dz e^{2T}\right]-4\La e^{2(DA+\Ds S+\Dd T)}=0$$
$$
\frac{\de S_{\mt{qm}}}{\de S}=0\ \Ra \ 2D\Ds\ddot{A} +2\Ds(\Ds-1)\ddot{S}+2\Dd(\Ds-1)\ddot{T}$$
$$+\2 (\Ds-1)e^{2(DA+(\Ds-1)S+\Dd T)}\left[2\Dz +\Dd ke^{-2T}-\2\Dz e^{2T}\right]-4\La\Ds e^{2(DA+\Ds S+\Dd T)}=0$$
$$
\frac{\de S_{\mt{qm}}}{\de T}=0\ \Ra \ 2D\Dd\ddot{A} +2(\Ds-1)\Dd\ddot{S}+2\Dd(\Dd-1)\ddot{T}$$
$$+\4 e^{2(DA+(\Ds-1)S+\Dd T)}\left[4\Dz\Dd +2\Dd(\Dd-1) ke^{-2T}-\Dz(\Dd+1) e^{2T}\right]-4\La\Dd e^{2(DA+\Ds S+\Dd T)}=0$$
It should be mentioned that there is a fourth equation which can for example be derived by varying $W(t)$ in the un-gauge fixed action. However, as is usual in general relativity it is not linearly independent, although it can constrain the initial conditions. A simple rearrangement of the equations gives us
\be
\ddot{A}-\frac{4\La}{\Dh-2} e^{2(DA+\Ds S+\Dd T)}=0
\label{eq:4A}
\ee
\be
\ddot{S}+e^{2DA}\left\{\2 e^{2(\Ds-1)S}\left[ e^{2\Dd T}-\2 e^{2(\Dd+1)T}\right]-\frac{4\La}{\Dh-2} e^{2\Ds S}e^{2\Dd T}\right\}=0
\label{eq:4S}
\ee
and
\be
\ddot{T}+\2 e^{2(DA+(\Ds-1)S)}\left[\2 ke^{2(\Dd-1)T}- e^{2\Dd T}+\4 (1+\textstyle{\Ds\over\Dd}) e^{2(\Dd+1)T}\right]=0
\label{eq:4T}
\ee
\vspace{5mm}
\\ 
{\bf  Solutions:}
One can immediately find the vacuum solutions, i.e. when the internal manifold is frozen. For constant $S$ and $T$, $\ddot{S}=\ddot{T}=0$ and from (\ref{eq:4S}) and (\ref{eq:4T}) we have
\be
 ke^{-2T}- 2+\2 (1+\frac{\Ds}{\Dd}) e^{2T}=0
\label{eq:4soln}
\ee
and
\be
\2 \left[ 1-\2 e^{2T}\right]-\frac{4\La}{\Dh-2} e^{2 S}=0
\ee
Substituting $k$ in (\ref{eq:4soln}) we have
$$
\left(1-\frac{\Dz}{2\Dd}\right)e^{-2T}- 2+ \left(1+\frac{\Dz}{2\Dd}\right) e^{2T}=0
$$
This has two solutions
\be
e^{2T}=1,\ \frac{2\Dd-\Dz}{2\Dd+\Dz}
\ee
and correspondingly
\be
e^{2S}=\frac{\Dh-2}{16\La},\ \left(\frac{\Dh-2}{16\La}\right)\left(\frac{2\Dd+3\Dz}{2\Dd+\Dz}\right)
\ee
Indeed these are the right vacuum solutions for the full higher dimensional Einstein's equations. The first one corresponds to the symmetric case, while the second one to the squashed case. 
 
 One can now imagine a situation where the universe started out in a symmetric  phase (say, that corresponds to the minima of the effective potential for the squashing field) starts to roll over or tunnel through the potential barrier (the maxima perhaps corresponding to the squashed vacuum) due to classical or quantum fluctuations/excitations and this may be accompanied by inflation. Then it can either settle to another minima or continue to evolve as a quintessence field. Indeed associated with the squashed vacuum solution one finds an exponential inflationary growth of the external universe. In the  $W=0$ gauge 
\be
e^{2A(t)}=e^{\Gamma_{\mathtt{I}} t} \mx{ with exponent } \Gamma_{\mathtt{I}}=\sqrt{\frac{8\La}{D(\Dh-2)}}
\ee
This solution is none other than the deSitter vacuum $dS_{D+1}\otimes G_{\mt{sq}}$, which in our dynamic universe model is just a phase\footnote{Indeed one does not expect the internal manifold to stay at the unstable squashed vacuum, but to slowly roll over but provided the slow roll conditions \cite{slow} are satisfied at the top of the hill we can still get inflation.}. 

Let us now see whether our toy model of pure Kaluza-Klein theory can also provide a ``quintessential solution'' of an accelerating universe where $T$ is say rolling towards $-\infty$. In this phase of evolution  the smallest exponent in the effective potential $V_{\mt{qm}}$ dominates, $T$ essentially rolling down  $e^{2(\Dd-1)T}$. Thus we can ignore all the other terms in the effective action (\ref{eq:4qe-action}-\ref{eq:4q-potential}). This effectively conceals the curvature of $G/H$ and we are left with the product space $M_{D+1}\otimes (G/H)_{\Dz}\otimes H$, where $H$ acquires an internal curvature $(1/4)k$ while $G/H$ becomes flat. It is easy to find a deSitter type solution:
\be
S(t)=A(t) \mx{ and } T(t)=T_0-A(t)
\label{eq:4Q-soln}
\ee
 In the $W=0$ gauge 
\be
e^{2A(t)}=e^{\Gamma_{\mathtt{q}} t} \mx{ with exponent } \Gamma_{\mathtt{q}}=\sqrt{\frac{8\La}{(D+\Dz)(\Dh-2)}}
\label{eq:4exponent}
\ee
We notice that the quintessence exponent is smaller than the inflation exponent as it should be. 

The above solutions certainly suggest that it may be possible to realize an inflationary or quintessence paradigm using the squashing field of the internal manifold, and merit further investigation. Among other things, one has to firstly perform a slow roll analysis of the potential for inflation. Subsequently, one should incorporate matter-radiation into the picture to study reheating, quintessence etc. For one, this would  no doubt ameliorate the exponential inflation and quintessence to a more standard power law type. Finally, one needs to account for stabilizing effects like brane gas \cite{brane,easson,kaya} or fluxes \cite{moduli}, because typically in extra dimensional cosmology, the size moduli is unstable and tends to expand\footnote{This can also be seen from our quintessence solution (\ref{eq:4exponent}), $S(t)$ is expanding. This is also related to the issue of  dilaton stabilization in string theory, the dilaton being the radion corresponding to the circular compactification of the 11 dimensional supergravity/M-theory.}. In the next section we try to address some of these issues.
\setcounter{equation}{0}
\section{{\bf COSMOLOGICAL SCENARIOS }}
{\bf Partial Stabilization and Conformal Transformation:} We notice that in the quintessence solution (\ref{eq:4exponent}) a combination of the size and the shape moduli, viz. $S+T$ remains constant and this phenomenon of a linear combination of  fields becoming frozen is recurrent  in  dynamics with many scalar fields specifically with exponential potentials \cite{Anupam}. Recently it has also come to attention that several other  mechanisms involving brane gas and fluxes can also stabilize the moduli at least partially. Indeed  stabilization of moduli is an intriguing and complicated problem which arises in almost all modern unified theories like String/M-theory and ideally the stabilization mechanisms should be included before we study cosmology with the moduli fields. However here we take a short cut and assume that the moduli is,  say partially stabilized: 
\be
\sigma=\psi+\e\te=\mt{ constant} \label{moduli} 
\label{eq:stabilization}
\ee 
 at least approximately with may be $\e$ varying slowly between different
cosmological eras. If the stabilization is achieved at a much higher scale or if it is dynamically stabilized as noted above, the corrections to the field theory potential (\ref{eq:4effpot}) can be ignored. Indeed different mechanisms may be at work at different cosmic times but as a first approximation we assume $\e$ to be a constant. Technically this assumption simplifies the analysis greatly as we are left with just a single scalar field potential which has been studied
extensively both in the context of inflation and quintessence. 

Before we implement (\ref{eq:stabilization}) it is useful to perform the following conformal rescalings in the field theory action (\ref{eq:4e-action}) that we have derived earlier:
\be
e_a{}^m=\Delta e'_a{}^m\ ;\ \Phi=(\Delta)^{-1}\Phi'\ ;\ \Delta=\Phi'^{\frac{\Ds}{\Dh-2}} \Te'^{\frac{\Dd}{\Dh-2}} \,,
\ee
This leads to the action 
\be
S=\frac{1}{16\pi G_{D+1}}\int d^{D+1}x\ e^{-1}[R+K-V],
\ee
where\footnote{We have re-introduced the $k$'s explicitly so that we can  look at the more general case when $H$ can also be a product of $U(1)$'s because, as it will soon become clear, physically it is  quite interesting.}
\beqy
K=-\frac{1}{\Dh-2}\biggl[\Ds(D-1)(\p\psi)^2+\Dd(\Dz+D-1)(\p\te)^2+2\Dd(D-1)\p_{a}\psi \p^{a}\te\,\biggr]\,\\
V=2\widehat{\La} e^{-\frac{2(\Ds\psi+\Dd\te)}{\Dh-2}}-\frac{1}{4}\biggl[(\Dz+2(1-k)\Dd) e^{-2\psi}-\Dd (1-k) e^{-2(\psi-\te)}+k\Dd e^{-2(\psi+\te)}\biggr]\,
\eeqy
Now using eq.(\ref{moduli}), the kinetic term becomes
\be
K=-A^2(\p\te)^2\,,
\ee
where
\beqy
A^2=\frac{1}{\Dh-2}\biggl[\e^2\Ds(D-1)&-&2\e\Dd(D-1)+\Dd(\Dz+D-1)\biggr]\,,
\eeqy
and the potential term becomes
\beqy
V=M_*^2\biggl[\widetilde{\La} e^{\frac{2(\e\Ds-\Dd)\te}{\Dh-2}}&-&\frac{1}{4}\biggl[(\Dz+2\Dd(1-k)) e^{2\e\te}-\Dd(1-k) e^{2(\e+1)\te}+k\Dd e^{2(\e-1)\te}\biggr]\biggr]\nonumber\\
\equiv M_*^2\widetilde{V}(\te)\,,\label{potential}
\eeqy
where
\be
M_*\equiv e^{-\sigma}\ ;\ \widetilde{\La}\equiv 2\widehat{\La}e^{2\frac{D-1}{\Dh-2}\sigma}
\ee

As is clear from eq.(\ref{potential}), depending upon the values of $\e$, $k$ and
$\widetilde{\Lambda}$, numerous cosmological scenarios can emerge. Pending an exhaustive
analysis of all of them, we focus below on only some of the most interesting cases.

\vskip 0.1in
\noindent
{\bf Quintessential Inflation:} Consider when $1+\Dz/2 >\e\geq 1$,
$k=0$ ($SU(2)$ to $U(1)$ being a typical example) and
$\widetilde{\Lambda}>0$, the potential (eq.~(\ref{potential})) for which is shown in
figure 1. As is clear this may realize a quintessential-inflation
scenario~\cite{peebles}: the shape field starts to roll down (or may tunnel through
the barrier) from the potential hill\footnote{ One can start the evolution either from near the  potential barrier or from the ``flattish'' potential minimum by giving a small initial kick which can be imagined to arise from classical or quantum fluctuations.} near $\te=0$
corresponding to a symmetric state ($G_L$ isometry) of the internal
manifold. This stage is accompanied by inflation although some fine tuning of the parameters is necessary to satisfy the slow roll conditions 
\be
\epsilon_H \equiv 3\frac{\dot{\te}^2}{2V+\dot{\te}^2}\ll 1
\ee 
and
\be
\eta_H\equiv -\frac{\ddot{\te}}{H\dot{\te}}\ll 1
\ee 
For several cases we can obtain a reasonable number of e-foldings, around 50-60, which is generally required to solve the cosmological flatness and horizon problems.  For e.g. for $\Ds=8$, $\Dd=1$ ($SU(3)\ra U(1)$) and the parameters $\epsilon=1$ and $\widetilde{\Lambda}=2.49$, we could
obtain around 60 e-foldings\footnote{ We checked that $\epsilon_H< 0.1$ during the inflationary phase, although $\eta_H$ is a little high, $\leq 0.5$ in this particular case.}. Most of the e-foldings are obtained as the field rolls through the maximum, indicating that the potential is not generically flat for the entire parameter range. For a given value of $\e$ the curvature at the maximum is determined by  the parameter $\Lambda$, which has to be somewhat fine-tuned (one part in hundred) to obtain a sufficiently flat potential between the minima and the maxima. Once the parameters have been so chosen no further fine-tuning of the initial conditions are necessary. For example we find that we may start from the minimum with an initial velocity within a range $[v_{\mt{min}},v_{\mt{min}}+\Delta v]$ and consistently obtain  a large number of  e-foldings. Even when $\Delta v/v_{\mt{min}}\sim 1$  the number of e-foldings only diminishes by half. 
\begin{figure}[!h]
\begin{center}
\includegraphics[scale=0.45,angle=270]{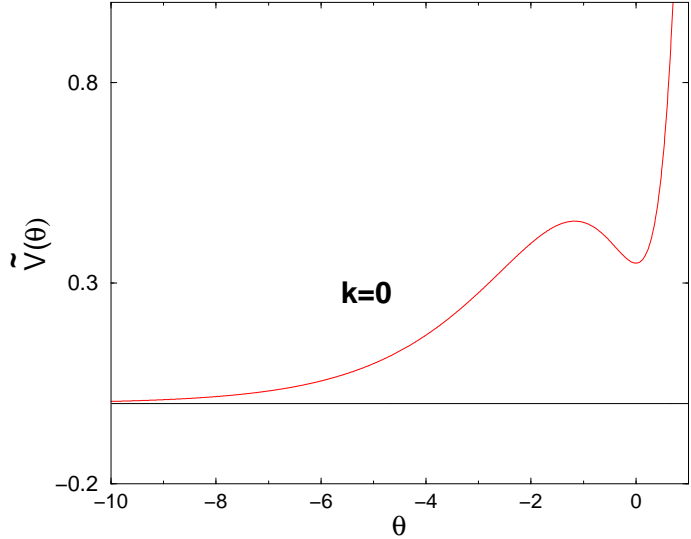}
\end{center}
\caption{The potential $\widetilde{V}(\theta)$ for $k=0$.}
\label{figps1}
\end{figure} 
\vskip 0.1in

In Planckian units inflation ends around $\te=-2$ when $\epsilon_H$ becomes large and a significant portion of the potential energy has been converted into kinetic energy ($\dot{\te}^2\sim V(\te)$); the universe  enters a deflationary or kinesis phase~\cite{spokoiny,peebles}. Matter/radiation entropy can be produced through gravitational particle production as discussed in~\cite{ford,peebles}. This is presumably closely followed by radiation and then a matter dominated era. These aspects of the evolution will be studied elsewhere. Here, we point out that previously, quintessential models with exponential potentials have been studied~\cite{copeland} and the scalar field slowly rolling down the flat exponential potential is known to possess scaling solutions where the evolution of the scalar field essentially  mimics that of the barotropic fluid, allowing the (external) universe to evolve as it would in the ordinary radiation/matter dominated era. However, it is possible for the potential energy of the quintessence field to start dominating the matter-energy content of the universe at a later point, leading to a second phase of inflation from which the universe never recovers. Indeed, this late inflationary phase has been ascribed to the small cosmological constant that we observe today eq.(\ref{cosmos})~\cite{exp}.

In our model it is actually quite non-trivial to be able to generate the hierarchy between the Plank mass and the cosmological constant as stringent constraints arise from its connection to particle physics as we will exemplify below. If $\te_c$ is the current value of the squashing parameter, then we find that
\be
\lambda \sim M_p^2~M_*^2~\widetilde{V}(\te_c)\,,\label{lambda}
\ee
However, $M_*$ is not arbitrary but instead fixed by particle physics. In the squashed internal manifold, the gauge field action coming from the Kaluza-Klein reduction looks like~\cite{TB}
\beqy
S_{\mt{gauge}}=\frac{M_p^2}{16\pi e^{2(\psi+\te)}}\int dx^{D+1}[g^K_{\as\bs}F^{\as}_{mn}F^{\bs mn}
+M_{\as\bs}A_m^{\az}A^{\bz m}+\dots ]\, \label{gauge}
\eeqy
where the mass-matrix for the broken gauge bosons (corresponding to the isometries along $G/H$ directions) are given by
\be
M_{\as\bs}=e^{-2\psi}[e^{2\te}+e^{-2\te}-2]\,.
\ee
From eq.(\ref{gauge}) it is clear that  we have in general a time varying  fine structure constant whose value today is given by
\be
\al=4\frac{M_*^2}{M_p^2}e^{2(\e-1)\te_c}\,.\label{fine}
\ee
The time variation of $\al$ depends on both how the quintessence field varies, which is slow over cosmological scales, but also on the coupling exponent $\e-1$. To make  precise statements one has to evolve the squashing field numerically or even analytically as exact solutions are known to exist for exponential potentials~\cite{copeland,Anupam}. However we will find out in the last subsection that essentially requiring consistency with the quintessence cosmology along with the various observational bounds coming from time variation of $\al$ and fifth force experiments constrain $\e-1$ to be very small ($<10^{-3}$).

 From an inspection of eq.(\ref{lambda}) and eq.(\ref{fine}) we find  that we now have two parameters, $\te_c$ and $M_*$ to fit two values $\la$ and $\al$, one coming from astrophysics and the other from particle physics  respectively. Can this  be accomplished without any fine tuning? From 
(\ref{fine}) and (\ref{lambda}) one finds
\be
\lambda \sim \frac{\al}{4}e^{-2(\e-1)\te_c}M_p^4~\widetilde{V}(\te_c)\,,\label{lambda2}
\ee
Now, if $k\neq 0$, this gives us a large constant term
\be
\lambda \sim -\frac{\al}{16}M_p^4k\Dd+\dots\,,
\ee
Thus although cases like $k>0$ with $\e > 1+\Dz/2$ has a potential which looks very similar to the case we are discussing here, one cannot get quintessence out of it without addressing the naturalness (or fine tuning) issue. If  $k=0$ as in our case,  we can indeed generate the hierarchy without fine tuning. In particular, for the example chosen, $\te_c=-60$ and $M_*=0.04~M_p$ gives us eq.(\ref{cosmos}) and $\al\sim 1/150$. Further, we observe 
\be
M_{\as\bs}=\frac{\al}{4}M_p^2(1-e^{2\te_c})^2\,.\label{mass}
\ee
When $\te_c\ll-1$ we have the mass of the broken gauge bosons $M\sim 10^{-2}M_p$.
Thus this mechanism would naturally explain gauge symmetry breaking in (S)GUT theories.

\vskip 0.1in

\noindent
{\bf Double-well Inflation:} Next let us  look at the case when symmetry breaking takes place via the usual Higgs-like mechanism. Consider the case when again the $\e$ parameter lies in the range $1+\Dz/2>\e>1$ with $\widetilde{\Lambda}>0$, but $k>0$. The potential looks like a double well just like in the ordinary Higgs mechanism (the solid line in figure 2a), also typically suited for inflationary cosmology. The symmetric minima is located around zero but the non-symmetric vacuum is the global minima situated away from zero in the negative axis. Again for certain parameters the slow roll conditions are satisfied around the potential barrier so that the phase transition can be accompanied by inflation.  To be specific let us look at the example when $SU(3)$ is broken to $SO(3)$ ($\Ds=8$ and $\Dd=3$). For $\e=1.1$ and $\widetilde{\La}=2.64$ the slow roll conditions are satisfied and we get 50 e-foldings~\footnote{We checked that $\epsilon_H< 0.1$ and $\eta_H<0.2$ during the inflationary phase.}. The fine-tuning estimates are similar to that of the quintessential inflation case.  For the choice of parameters above, the true minimum (asymmetric vacuum) is located around  $\te_{\mt{min}}\equiv\te_c=-1.7$. Then, using $\al=1/132$ we obtain  $\ M_*\sim .07~M_p$.
 For the values mentioned, we find $M\sim 10^{-2}M_p$.
 
 Indeed this scenario cannot solve the cosmological constant problem but serves as a regular inflationary scenario. An interesting study would be to consider a double phase transition where the initial isometry group $G$ is first broken to $H$ as explained here which gives rise to inflation and then the second phase transition $H\ra K$ (where $K\subset H$) can account for quintessence in much the same manner as the quintessential-inflation case. To make matters concrete one could have $SU(3)\ra SO(3)\ra SO(2)\sim U(1)$ rather than directly going to $U(1)$ as discussed in the earlier subsection. Note, that the latter phase transition $SO(3)\ra SO(2)$ also corresponds to $k=0$ and hence a quintessence scenario is feasible.
\vskip 0.1in
\begin{figure}[!h]
\begin{center}
\includegraphics[scale=0.45,angle=270]{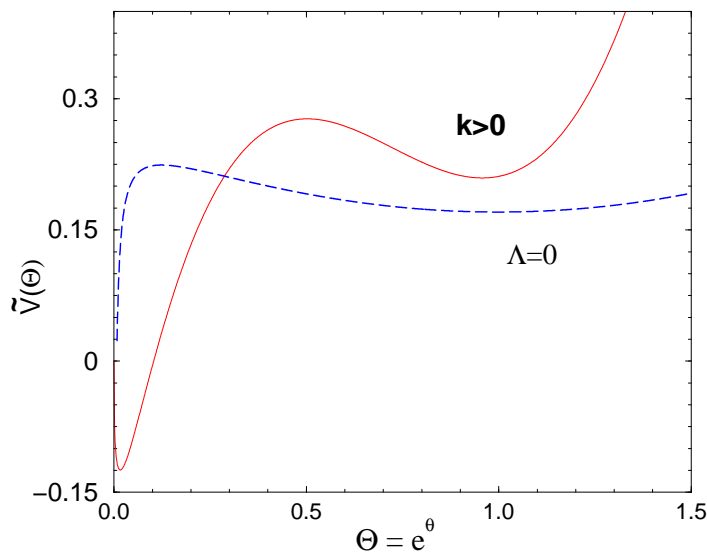}
\end{center}
\caption{The potential $\widetilde{V}(\Theta)$ for a) double-well inflation (solid line) and b) quintessence ($\tilde{\Lambda}=0$) (dashed line).}
\label{figps2}
\end{figure} 
\vskip 0.1in
\noindent
{\bf Quintessence:} We have so far seen two different ways that the $\te$ evolution can break symmetry. Now we consider an intriguing possibility of an opposite process, viz. symmetry restoration being accomplished by the squashing field. For the special case when $\widehat{\La}=0$ our potential eq.(\ref{potential}) looks like 
\be
\widetilde{V}=-\frac{1}{4}\left[2\Dz e^{\sqrt{32\pi}\frac{\e}{A}\te}-\frac{1}{2}\Dz e^{\sqrt{32\pi}\frac{\e+1}{A}\te}+k\Dd e^{\sqrt{32\pi}\frac{\e-1}{A}\te}\right]
\label{eq:qpotential}
\ee
Here we have redefined $\te\ra (A/\sqrt{8\pi})\te$ to make $K$ have the canonical form. The potential is qualitatively sketched in the dashed line of figure 2b. From the form of the potential  it is clear that the picture resembles the scenario considered in \cite{bshape}  where the quintessence field (squashing field) is trapped at the top of the potential hill until recently. This is possible due to the friction provided by the expansion of the universe, as argued in \cite{bshape}. As the universe expands the friction due to the Hubble parameter drops and eventually the squashing field may start rolling slowly towards the symmetric minima acting as a quintessence field.

It is also interesting to note that (\ref{eq:qpotential}) resembles a class of potentials considered in ~\cite{alpha}. Motivated by the success of the Albrecht-Skordis potential in describing a Q-field evolution \cite{skordis} consistent with Big Bang Nucleosynthesis (BBN) and  quintessence domination at the current epoch, the authors in~\cite{alpha} considered a sum of three exponential potentials which recovers the Albrecht-Skordis potential as a limiting case. In particular, it was examined if the evolving quintessence field could account for the time variation of the fine structure constant, with reasonable success. It was also remarked that such a potential could arise from the moduli of the internal manifold which is indeed corroborated by our results.   We leave it as a future exercise to make the connection between the evolution of the squashing field and quintessence/$\al$-variation more precise in these contexts.  \\
{\bf Fifth Force, $\al$ Variation and CMB Fluctuations:} We have seen so far that the effective  potential for the squashing field that one obtains from higher dimensional theories may be able realize some of the basic features of inflationary and quintessential cosmologies. Constructing realistic scenarios would however envisage testing these models against various other observational bounds, most notably  coming from CMB fluctuations, data on time variation of fine structure constant  and fifth force experiments. Here we try to provide  approximate estimates of these effects. 

First let us look at the quintessential-inflation scenario. The low scale of the cosmological constant implies that the quintessence field is effectively massless and therefore mediates a  ``fifth force''. Various null experiments on fifth force essentially put bounds on the coupling exponent of the scalar field to the electro-magnetic $F^2$ term:
\be
|\nu|< 10^{-3} \mx{ where }S_F=-\frac{e^{\nu\te}}{16\pi\al}F^2
\ee
Typically this places a bound on the $\e$ parameter
\be
|\e-1|<10^{-3} \label{epsbound}
\ee
This suggests the importance of finding a stabilization mechanism with $\e$ equal or very close to one\footnote{In our example we chose $\e=1$.}. Observations on $\al$ variation (see ref.~\cite{alpha} for details) also seem to indicate similar bounds. In our model
\be
|\frac{\dot{\al}}{\al}|=|\nu|\dot{\te}<10^{-15}yr^{-1}
\ee
Since at late times only the last exponent dominates, the squashing field essentially rolls along an exponential potential. For such attractor solutions it is known that the kinetic energy is a fraction of the scalar potential energy which is identified with the effective cosmological constant. This gives us a handle to estimate the $\te$ variation
\be
\dot{\te}\approx 10^{-12} yr^{-1}
\ee
which again implies~(\ref{epsbound}). Finally one can also look at observational bounds coming from variation of $\al$ on cosmic scales. Assuming  $\frac{\Delta\al}{\al}$ to be small one finds
\be
\frac{\Delta\al}{\al}\approx -\nu\Delta\te
\ee
To match all the  bounds~\cite{alpha} coming from observations at different epochs one has to perform  numerical simulations with appropriate radiation-matter density as for example was done in [?]. However, one can quickly estimate the expected variation in our model since BBN. We have seen that typically to be consistent with quintessence cosmology $\te_c\sim -50$ while inflation ends around $\te \sim -2$, so that  $(\te_{\mt{BBN}}-\te_c)\sim 50$. For BBN it is known that $\frac{\Delta\al}{\al}\sim 10^{-2}$ which again seem to suggest a bound on  $\e$ similar to~(\ref{epsbound})!

Let us now try to estimate the CMB fluctuations arising from the inflationary scenarios in our model. As we noted earlier, most of the e-foldings come from the flattish maxima in our potential. The spectral tilt and the amplitude of the CMB  fluctuations would then naturally originate at the viscinity of the maximum. In the approximation that the maximum is generated by two competing expotentials of the form
\be
V=M_{\mt{r}}^4(V_1e^{-\al_1\te}-V_2 e^{-\al_2\te})
\ee
where $M_{\mt{r}}$ is the reduced Planck mass and 
\be
\al_2>\al_1
\ee
we find
\be
V_{\mt{max}}=M_{\mt{r}}^4V_2\left(\frac{\al_2-\al_1}{\al_1}\right)\left(\frac{\al_1V_1}{\al_2V_2}\right)^{\frac{\al_2}{\al_2-\al_1}} \label{amplitude}
\ee
From~(\ref{amplitude}), it is clear that if $\al_1$ is close to $\al_2$ then $\al_2/(\al_2-\al_1)$ can easily be a large number and a small hierarchy between $V_1$ and $V_2$ can create a large hierarchy\footnote{Typically in our model we have $\al_2/(\al_2-\al_1)\sim 10$ so that even a modest  $V_1/V_2\sim 0.1$ will be able to achieve the required hierarchy.}  sufficient to explain the amplitude of CMB fluctuations. Approximately we have
\be
\de_H\sim\frac{60\sqrt{V_{\mt{max}}}}{M_{\mt{r}}^2(\Delta\te)_{\mt{infl}}}
\ee
so that
$$
V_{\mt{max}}\sim 10^{-12}\Ra \de_H \sim 10^{-5}
$$
One could also compute the spectral tilt from the formula
\be
n_S=1-6\e+2\eta
\ee 
which in our case implies
\be
n_S=1-2\al_1\al_2(1-(\al_1+\al_2)(\te_{50}-\te_{\mt{max}})) \label{spectral}
\ee
where $\te_{50}$ is the point from which approximately 50-60 e-foldings ensure. Eq.~(\ref{spectral}) tells us that for a range\footnote{In our specific example in the quintessential-inflation scenario we found that $\te_{50}$ had to lie within $0.42\pm 0.01$ while for the example of pure inflation  $\te_{50}$ had to be within $0.49\pm 0.01$.} of $\te_{50}$ one may be able to explain the observed spectral tilt $n_S=0.99\pm.04$~\cite{Bennett03}. Of course matching the spectral tilt and the amplitude of CMB fluctuations with  the required 50-60 efoldings will require fine-tuning of the parameters as well as restrictions on the group theoretic parameters like the dimensions of the group and the subgroup manifold and it should be interesting to exlore these aspects in further details. 

\vskip 0.1in
\section{{\bf SUMMARY AND FUTURE RESEARCH}} 
In \cite{TB}, we had tried to explain how a dynamical internal manifold can break gauge symmetry partially. In particular, a transition of the internal manifold from a symmetric vacuum to a squashed, and hence less symmetric vacuum was suggested. In this paper we have tried to analyze in more detail when and how such a transition can occur along with its cosmological implications. For simplicity, we focused on the case when the internal manifold is a simple Lie group $G$ and we are interested in breaking the isometry group from $G_R \ra H_R$. Further, we assumed $H$ to be either simple or a product of $U(1)$'s. We first studied the dynamics using collective coordinates characterizing the size (radius) and the shape (squashing parameter)  of the internal manifold. We derived an effective potential for the squashing parameter which gives nice Newton's law type equations of motion which is useful to obtain exact or approximate cosmological solutions. In particular we obtained solutions for early and late times which resembles inflation and quintessence respectively. However, to make things concrete one has to perform a more detailed analysis.

Accordingly, we first obtained the effective potential by performing dimensional reduction and then conformal re-scalings of the higher dimensional gravitational action. To proceed further we assumed that the moduli is partially stabilized which gave us a potential for  the squashing field as a sum of four exponentials. This leads to numerous different and interesting  cosmological scenarios, specially with regards to quintessence and inflation, of which we considered three specifically. 

In the first two cases that we study one can imagine that the squashing field is initially  trapped in a flat  potential well (symmetric state). Quantum or classical fluctuations can then instigate  a symmetry breaking phase transition of the internal manifold with the rolling over phase of the squashing field across the potential barrier being accompanied by inflation. With some fine tuning of the parameters it is possible to satisfy the slow roll conditions and obtain around 50-60 e-foldings which is sufficient to solve the cosmological problems like the horizon and flatness problems. The fate of the inflaton however  differs in the two cases. In the first one the potential does not have a second mimina but rather rolls to zero asymptotically. This is essentially the scenario of quintessential inflation, where the inflaton at a later stage of evolution can account for a small effective cosmological constant. Indeed in the case that we consider we find that it is possible to generate a hierarchy between the Planck mass and the extremely small current vacuum energy density. The second scenario describes a more conventional Higgs-like symmetry breaking mechanism with double well inflation. What could be interesting is if one combines the two scenarios and consider a twin phase transition, where the first corresponds to the usual double well inflation whereas the second gives rise to late time quintessence. Firstly, this is realistic because it is believed that there were at least two symmetry breaking transitions (GUT and electro-weak) and secondly one would have a lot more parameters to play with in order to meet with experimental bounds coming from nucleosynthesis, density perturbations, quintessence equation of state etc. 

We also considered a third possibility where the symmetry could be restored, the quintessence field starting from an asymmetric state at the top of a potential hill and moving gradually towards the symmetric minimum. We also indicated its possible connection with cosmic variation of fine-structure constant.

Several issues remain open. For example, what happens when the fluxes are turned on? Can it really  stabilize  the moduli?  One could also  introduce warping and study the model in the context of a brane-world. This would ameliorate the fine tuning problems that exists in the double well type of potentials that we obtained for some parameter ranges. Finally,  one can try to analyze the dynamics of the internal manifold without assuming any prior stabilization mechanism. Thus, it may be that while the transition of the squashing field (in the double well case) generates inflation, it is the running of the radion that is responsible for quintessence, thereby eliminating the need to fine tune the initial potential minimum for the squashing field. It may also be a case of assisted inflation \cite{Anupam} where both the fields again become important.  

Finally, several other variations of the same idea can and should be studied for realistic phenomenological applications. Firstly, one can generalize the model from pure gravity to supergravity. The extra fields (like dilaton) may contain   scalars that are relevant to the squashing dynamics, while the form fields will give rise to additional potential terms. Secondly, it should be possible to generalize the internal manifolds and metrics that we considered in this paper which can change some of the parameters and even the nature of the effective potentials. For example it should  not be too difficult to generalize this mechanism to more complicated internal manifolds like the coset spaces (and in particular those which can give Standard Model like gauge groups), and at least in principle to some of the more interesting inhomogeneous spaces. 
\vspace{5mm}\\
{\bf Acknowledgements:} The authors thank Anupam Mazumdar and Horace Stoica for useful discussions and suggestions. This work is supported in part by the Natural Sciences and Engineering Research Council of Canada and in part by the Fonds Nature et Technologies of Quebec.

\end{document}